\documentclass{article}

\usepackage{amsfonts}
\usepackage{amssymb}
\usepackage{amsmath}
\usepackage{amsthm}
\usepackage{amscd}
\usepackage[margin=3.4cm]{geometry}

 \newtheorem{theorem}{Theorem}[section]
 \newtheorem{lemma}[theorem]{Lemma}
 \newtheorem{corollary}[theorem]{Corollary}
 \newtheorem{definition}[theorem]{Definition}

\theoremstyle{plain}

\def\id{{\hbox{\Bbb I}}}

\def\duzomniejsze{<\kern-.7mm<}
\def\duzowieksze{>\kern-.7mm>}

\def\textbf#1{{\bf #1}}
\def\beq{\begin{equation}}
\def\eeq{\end{equation}}
\def\be{\begin{equation}}
\def\ee{\end{equation}}
\def\ben{\begin{eqnarray}}
\def\een{\end{eqnarray}}
\def\beqa{\begin{eqnarray}}
\def\eeqa{\end{eqnarray}}
\def\eea{\end{array}}
\def\bea{\begin{array}}
\def\cP{{\cal P}}
\newcommand{\bei}{\begin{itemize}}
\newcommand{\eei}{\end{itemize}}
\newcommand{\bee}{\begin{enumerate}}
\newcommand{\eee}{\end{enumerate}}

\newcommand*{\beopt}{\begin{equation}}
\newcommand*{\eeopt}{\end{equation}}

\newcommand{\ket}[1]{|#1\rangle}                    
\newcommand{\bra}[1]{\langle#1|}                    
\newcommand{\proj}[1]{|#1\rangle\langle#1|}         

\newcommand{\cancel}[1]{}

\newcommand*{\cE}{\mathcal{E}}

\newcommand*{\cH}{\mathcal{H}}

\newcommand*{\cR}{\mathcal{R}}

\newcommand*{\acc}{\mathrm{acc}}
\newcommand*{\LOPC}{\mathrm{LOPC}}

\def\hcal{{\cal H}}

\def\pcal{{\cal P}}

\def\>{\rangle}
\def\<{\langle}

\def\tr{{\rm Tr}}

\def\ep{\epsilon}

\def\kd{K_D}

\def\id{I}
\def\dr{D_R}

\def\ot{\otimes}

\def\twointr{reduced intrinsic information }
\def\Twointr{Reduced intrinsic information }

\def\half{\frac{1}{2}}

\begin{document}
\title{Unifying classical and quantum key distillation}

\author{Matthias Christandl\thanks{Centre for Quantum Computation,
    University of Cambridge, United Kingdom, \newline {\tt
      \{m.christandl, jono, r.renner\}@damtp.cam.ac.uk}}
  \setcounter{mpfootnote}{\value{footnote}} \and Artur
  Ekert{$^\fnsymbol{footnote}$}\thanks{Department of Physics, National
    University of Singapore, Singapore, {\tt artur.ekert@qubit.org}}
  \and Micha\l{} Horodecki\thanks{Institute of Theoretical Physics and
    Astrophysics, University of Gda\'nsk, Poland, {\tt
      fizmh@univ.gda.pl}} \and Pawe\l{} Horodecki\thanks{Faculty of
    Applied Physics and Mathematics, Gda\'nsk University of
    Technology, Poland {\tt pawel@mif.pg.gda.pl}} \and
  \setcounter{footnote}{1} Jonathan Oppenheim{$^\fnsymbol{footnote}$}
  \and Renato Renner{$^\fnsymbol{footnote}$}}

\maketitle

\begin{abstract}
  Assume that two distant parties, Alice and Bob, as well as an
  adversary, Eve, have access to (quantum) systems prepared jointly
  according to a tripartite state $\rho_{ABE}$. In addition, Alice and
  Bob can use local operations and authenticated public classical
  communication.  Their goal is to establish a key which is unknown to
  Eve. We initiate the study of this scenario as a unification of two
  standard scenarios: (i) key distillation (agreement) from classical
  correlations and (ii) key distillation from pure tripartite quantum
  states.

  Firstly, we obtain generalisations of fundamental results related to
  scenarios~(i) and~(ii), including upper bounds on the key rate,
  i.e., the number of key bits that can be extracted per copy of
  $\rho_{A B E}$.  Moreover, based on an embedding of classical
  distributions into quantum states, we are able to find new
  connections between protocols and quantities in the standard
  scenarios~(i) and~(ii).

  Secondly, we study specific properties of key distillation
  protocols.  In particular, we show that every protocol that makes
  use of pre-shared key can be transformed into an equally efficient
  protocol which needs no pre-shared key. This result is of practical
  significance as it applies to quantum key distribution (QKD)
  protocols, but it also implies that the key rate cannot be locked
  with information on Eve's side.  Finally, we exhibit an arbitrarily
  large separation between the key rate in the standard setting where
  Eve is equipped with quantum memory and the key rate in a setting
  where Eve is only given classical memory.  This shows that
  assumptions on the nature of Eve's memory are important in order to
  determine the correct security threshold in QKD.
\end{abstract}

\section{Introduction}

Many cryptographic tasks such as message encryption or authentication
rely on \emph{secret keys},\footnote{In the sequel, we will use the
  term \emph{key} instead of \emph{secret key}.} i.e., random strings
only known to a restricted set of parties. In
\emph{information-theoretic cryptography}, where no assumptions on the
adversary's resources\footnote{In this context, the term
  \emph{resources} typically refers to computational power and memory
  space.} are made, distributing keys between distant parties is
impossible if only public classical communication channels are
available~\cite{Shanno49,Maurer93}.  However, this situation changes
dramatically if the parties have access to additional devices such as
noisy channels (where also a wiretapper is subject to noise), a noisy
source of randomness, a quantum channel, or a pre-shared quantum
state.  As shown
in~\cite{Maurer93,CsiszarKorner,AhlCsi93,BB84,Ekert91}, these devices
allow the secure distribution of keys.\footnote{In certain scenarios,
  including the one studied in this paper, an authentic classical
  channel is needed in addition.}

This work is concerned with information-theoretic key distillation
from pre-distributed noisy data. More precisely, we consider a
situation where two distant parties, \emph{Alice} and \emph{Bob}, have
access to (not necessarily perfectly) correlated pieces of (classical
or quantum) information, which might be partially known to an
adversary, \emph{Eve}. The goal of Alice and Bob is to \emph{distill}
virtually perfect key bits from these data, using only an authentic
(but otherwise insecure) classical communication channel.

Generally speaking, key distillation is possible whenever Alice and
Bob's data are sufficiently correlated and, at the same time, Eve's
uncertainty on these data is sufficiently large. It is one of the
goals of this paper to exhibit the properties pre-shared data must
have in order to allow key distillation.

In practical applications, the pre-distributed data might be obtained
from realistic physical devices such as noisy (classical or quantum)
channels or other sources of randomness. Eve's uncertainty on Alice
and Bob's data might then be imposed by inevitable noise in the
devices due to thermodynamic or quantum effects.

\emph{Quantum key distribution (QKD)} can be seen as a special case of
key distillation where the pre-shared data is generated using a
quantum channel.  The laws of quantum physics imply that the random
values held by one party, say Alice, cannot at the same time be
correlated with Bob and Eve. Hence, whenever Alice and Bob's values
are strongly correlated (which can be checked easily) then Eve's
uncertainty about them must inevitably (by the laws of quantum
mechanics) be large, hence, Alice and Bob can distil key. Because of
this close relation between key distillation and QKD, many of the
results we give here will have direct implications to QKD.

Furthermore, the theory of key distillation has nice parallels with
the theory of \emph{entanglement distillation}, where the goal is to
distil maximally entangled states (also called \emph{singlets}) from
(a sequence of) bipartite quantum states.  In fact, the two scenarios
have many properties in common.  For example, there is a gap between
the \emph{key rate} (i.e., the amount of key that can be distilled
from some given noisy data) and the \emph{key cost} (the amount of key
that is needed to simulate the noisy data, using only public classical
communication)~\cite{RenWol03}. This gap can be seen as the classical
analogue of a gap between \emph{distillable entanglement} (the amount
of singlets that can be distilled from a given bipartite quantum
state) and \emph{entanglement cost} (the amount of singlets needed to
generate the state).

\subsection{Related work}

The first and basic instance of an information-theoretic key agreement
scenario is Wyner's wiretap channel~\cite{Wyner75}. Here, Alice can
send information via a noisy classical channel to Bob. Eve, the
eavesdropper, has access to a degraded version of Bob's information.
Wyner has calculated the rate at which key generation is possible if
only Alice is allowed to send public classical messages to Bob.
Wyner's work has later been generalised by Csisz{\'a}r and K{\"o}rner,
relaxing the restrictions on the type of information given to
Eve~\cite{CsiszarKorner}.  Based on these ideas, Maurer and Ahlswede
and Csisz{\'a}r have proposed an extended scenario where key is
distilled from arbitrary correlated classical information (specified
by a tripartite probability distribution)~\cite{Maurer93,AhlCsi93}.
In particular, Maurer has shown that two-way communication can lead to
a strictly positive key rate even though the key rate in the one-way
communication scenario might be zero~\cite{Maurer93}.

In parallel to this development quantum cryptography emerged: in 1984
Bennett and Brassard devised a QKD scheme in which quantum channels
could be employed in order to generate a secure key without the need
to put a restriction on the eavesdropper~\cite{BB84}. In 1991, Ekert
discovered that quantum cryptographic schemes could be based on
entanglement, that is, on quantum correlations that are strictly
stronger than classical correlations~\cite{Ekert91}. Clearly, this is
key distillation from quantum information.

The first to spot a relation between the classical and the quantum
development were Gisin and Wolf; in analogy to \emph{bound
  entanglement} in quantum information theory, they conjectured the
existence of \emph{bound information}, namely classical correlation
that can only be created from key but from which no key can be
distilled~\cite{GisWol00}. Their conjecture remains unsolved, but has
stimulated the community in search for an answer.


To derive lower bounds on the key rate, we will make repeated use of
results by Devetak and Winter, who derived a bound on the key rate if
the tripartite quantum information consists of many identical and
mutually independent pieces, and by Renner and K{\"o}nig, who derived
privacy amplification results which also hold if this independence
condition is not satisfied~\cite{DevWin04,RenKoe05}.


\subsection{Contributions}

We initiate the study of a unified key distillation scenario, which
includes key distillation from pre-shared \emph{classical} and
\emph{quantum} data (Section~\ref{section-unifying}). We then derive a
variety of quantitative statements related to this scenario. These
unify and extend results from both the quantum and classical world.

There are numerous upper bounds available in the specific scenarios
and it is our aim to provide the bigger picture that will put order
into this zoo by employing the concept of a \emph{secrecy monotone},
i.e., a function that decreases under local operations and public
communication (Section~\ref{section-bounds}), as introduced
in~\cite{CeMaSc02}.  The upper bounds can then roughly be subdivided
into two categories: (i)~the ones based on classical key
distillation~\cite{MauWol97c-intr} and (ii)~the ones based on quantum
communication or entanglement measures~\cite{Michal2001}.

The unified scenario that we develop does not stop at an evaluation of
the key rate but lets us investigate intricate connections between the
two extremes. We challenge the viewpoint of Gisin and Wolf who
highlight the relation between key distillation from classical
correlation and entanglement distillation from this very correlation
\emph{embedded} into quantum states~\cite{GisWol00}: we prove a
theorem that relates key distillation from certain classical
correlation and key (and not entanglement) distillation from their
embedded versions (Section~\ref{section-embedding}). This ties in with
recent work which established that key distillation can be possible
even from quantum states from which no entanglement can be
distilled~\cite{pptkey}.

A fruitful concept that permeates this work is the concept of
\emph{locking of classical information in quantum states}: let Alice
choose an $n$-bit string $x=x_1\ldots x_n$ with uniform probability
and let her either send the state $\ket{x_1}\ldots \ket{x_n}$ or the
state $H^{\otimes n}\ket{x_1}\ldots \ket{x_n}$ to Bob, where $H$ is
the Hadamard transformation. Not knowing if the string is sent in the
computational basis or in the Hadamard basis, it turns out that the
optimal measurement that Bob can do in order to maximise the mutual
information between the measurement outcome $y$ and Alice's string $x$
is with respect to a randomly chosen basis, in which case he will
obtain $I(X;Y)=\frac{n}{2}$. If, however, he has access to the single
bit which determines the basis, he will have $I(X;Y)=n$.  A
\emph{single bit} can therefore \emph{unlock} an arbitrary amount of
information.  This effect has been termed \emph{locking of classical
  information in quantum states} or simply \emph{locking} and was
first described in~\cite{DiVincenzo-locking}.  In this paper, we will
discuss various types of locking effects and highlight their
significance for the design and security of QKD protocols
(Section~\ref{sec:lock}).

Finally, we demonstrate that the amount of key that can be distilled
from given pre-shared data strongly depends on whether Eve is assumed
to store her information in a classical or in a quantum memory. This,
again, has direct consequences for the analysis of protocols in
quantum cryptography (Section~\ref{section-qkd}).

For a more detailed explanation of the contributions of this paper,
we refer to the introductory paragraphs of
Sections~\ref{section-bounds}--\ref{section-qkd}.

\section{The unified key distillation scenario}
\label{section-unifying}

In classical information-theoretic cryptography one considers the
problem of distilling key from correlated data specified by a
tripartite probability distribution $p_{ijk}$ ($p_{ijk}\geq 0$,
$\sum_{i,j,k} p_{ijk}=1$).  Alice and Bob who wish to distil the key
have access to $i$ and $j$, respectively, whereas the eavesdropper Eve
knows the value $k$ (see, e.g., \cite{MaurerWolf00CK}).  Typically, it
is assumed that many independently generated copies of the triples
$(i,j,k)$ are available\footnote{Using de Finetti's representation
  theorem, this assumption can be weakened to the assumption that the
  overall distribution of all triples is invariant under permutations
  (see~\cite{Renner05} for more details including a treatment of the
  quantum case).}.  The \emph{key rate} or \emph{distillable key} of a
distribution $p_{i j k}$ is the rate at which key bits can be obtained
per realisation of this distribution, if Alice and Bob are restricted
to local operations and public but authentic classical communication.

Before we continue to introduce the quantum version of the key
distillation scenario described above, let us quickly note that it
will be convenient to regard probability distributions as
\emph{classical states}, that is, given probabilities $p_i$, we
consider $ \rho=\sum_{i=1}^d p_i \proj{i}$, where $\ket{i}$ is an
orthonormal basis of a $d$-dimensional Hilbert space; we will assume
that $d<\infty$. In the sequel we will encounter not only classical or
quantum states, but also states that are distributed over several
systems which might be partly classical and partly quantum-mechanical.
To make this explicit, we say that a bipartite state $\rho_{AB}$ is
\emph{cq (classical-quantum)} if it is of the form $\rho_{AB}=\sum_i
p_i \proj{i}_A \otimes \rho^i_B$ for quantum states $\rho^i_B$ and a
probability distribution $p_i$. This definition easily extends to
three or more parties, for instance:
\begin{itemize}
\item a \emph{ccq (classical-classical-quantum) state} $\rho_{ABE}$
is of the form $\sum_{i,j} p_{ij} \proj{i}_A \otimes \proj{j}_B
\otimes \rho_E^{ij}$, where $p_{ij}$ is a probability distribution
and $\rho_E^{ij}$ are arbitrary quantum states.
\item the probability distribution $p_{ijk}$ corresponds to a
  \emph{ccc (classical-classical-classical) state} $\rho_{ABE} =
  \sum_{i,j,k} p_{ijk} \proj{ijk}_{ABE}$, where we use
  $\ket{ijk}_{ABE}$ as a short form for $\ket{i}_A\otimes \ket{j}_B
  \otimes \ket{k}_E$ (as above, the states $\ket{i}_A$ for different
  values of $i$, and likewise $\ket{j}_B$ and $\ket{k}_k$, are
  normalised and mutually orthogonal).
\end{itemize}

We will be concerned with key distillation from arbitrary tripartite
quantum states $\rho_{ABE}$ shared by Alice, Bob, and an adversary
Eve, assisted by {\it local quantum operations and public classical
  communication (LOPC)}~\cite{DevWin04,ChristandlR04-ABEkey,pptkey}. A
local quantum operation on Bob's side is of the form \beopt
\rho_{ABE} \mapsto (I_{AE}\ot \Lambda_B)(\rho_{ABE}) \ . \eeopt
Public classical communication from Alice to Bob can be modelled by
copying a local classical register, i.e., any state of the form
$\rho_{A A' B E}=\sum_i \rho_{A B E}^i\ot |i\>\<i|_{A'}$ is
transformed into $\rho'_{A A' B B' E E'}=\sum_i \rho_{A B E}^i \ot
|iii\>\<iii|_{A'B'E'}$.  Similarly, one can define these operations
with the roles of Alice and Bob interchanged.

The goal of a key distillation protocol is to transform copies of
tripartite states $\rho_{A B E}$ into a state which is close to \be
\tau^{\ell}_{ABE}={1\over 2^\ell}\sum_{i=1}^{2^\ell}|ii\>\<ii|_{AB}\ot
\tau_E \label{eq:ideal} \ee for some arbitrary $\tau_E$.
$\tau^\ell_{ABE}$ (also denoted $\tau^\ell$ for short) corresponds to
a perfect \emph{key of length $\ell$}, i.e., uniform randomness on an
alphabet of size $2^\ell$ shared by Alice and Bob and independent of
Eve's system.  We measure \emph{closeness} of two states $\rho$ and
$\sigma$ in terms of the trace norm $\| \rho-\sigma\|:=\half \tr
|\rho-\sigma|$.  The trace norm is the natural quantum analogue of the
variational distance to which it reduces if $\rho$ and $\sigma$ are
classical.

We will now give the formal definition of an LOPC protocol and of the
key rate.

\begin{definition}\label{def-LOPCprotocol}
  An \emph{LOPC protocol} $\cP$ is a family $\{ \Lambda_n\}_{n \in
    \mathbb{N}}$ of completely positive trace preserving (CPTP) maps
  \beopt \Lambda_n: (\hcal_A\ot\hcal_B\ot \hcal_E)^{\ot n}\to \hcal_A^n \ot
  \hcal_B^n \ot \hcal^n_E \eeopt
  which are defined by the concatenation of
  a finite number of local operation and public communication steps.
\end{definition}

\begin{definition}
\label{def:key-rate} We say that an LOPC protocol $\cP$ distills key
at rate $\cR_\cP$ if there exists a sequence $\{\ell_n\}_{n \in
  \mathbb{N}}$ such that
\begin{align} 
\limsup_{n \to \infty} {\ell_n\over  n} & = R_\pcal \\
\lim_{n \to \infty} \|\Lambda_n(\rho^{\ot n}_{ABE}) -
\tau^{\ell_n}_{ABE}\| & = 0 
\end{align}
where $\tau^{\ell_n}_{ABE}$ are the
ccq states defined by~\eqref{eq:ideal}. The \emph{key rate} or
\emph{distillable key} of a state $\rho_{ABE}$ is defined as
$K_D(\rho_{ABE}) :=\sup_{\cP} \cR_\cP$.
\end{definition}

The quantity $K_D$ obviously depends on the partition of the state
given as argument into the three parts controlled by Alice, Bob, and
Eve, respectively.  We thus indicate the assignment of subsystems by
semicolons if needed.  For instance, we write $\rho_{AD;B;E}$ if Alice
holds an additional system $D$.

As shown in Appendix~\ref{app:commlin}, the maximisation in the
definition of $K_D$ can be restricted to protocols whose communication
complexity grows at most linearly in the number of copies of
$\rho_{ABE}$.  Hence, if $d=\dim \cH_A \otimes \cH_B \otimes \cH_E
<\infty$ then the dimension of the output of the protocol is bounded
by $\log \dim \cH_A^n \otimes \cH_B^n \otimes \cH_E^n \leq c n \log d
$, for some constant $c$.



The above security criterion is (strictly) weaker than the one
proposed in~\cite{DevWin04}\footnote{The security criterion
  of~\cite{DevWin04} implies that, conditioned on \emph{any} value of
  the key, Eve's state is almost the same. In contrast, according to
  the above definition, Eve's state might be arbitrary for a small
  number of values of the key.}, hence $K_D(\rho_{ABE})$ evaluated on
cqq states is lower bounded by an expression derived
in~\cite{DevWin04}: \be K_D(\rho_{ABE}) \geq I(A:B)_\rho- I(A:E)_\rho
\ .
\label{eq:lowerbound} \ee This expression can be seen as a
quantum analogue of the well-known bound of Csisz\'ar, K\"orner, and
Maurer~\cite{CsiszarKorner,MaurerWolf00CK}.  Here $I(A:B)_\rho$
denotes the mutual information defined by $I(A:B)_\rho :=
S(A)_\rho+S(B)_\rho-S(AB)_\rho$ where $S(A)_\rho:=S(\rho^A)$ is the
von Neumann entropy of system $A$ (and similarly for $B$ and $E$). For
later reference we also define the \emph{conditional mutual
  information}
$I(A:B|E)_\rho:=S(AE)_\rho+S(BE)_\rho-S(ABE)_\rho-S(E)_\rho$.

Note also that the criterion for the quality of the distilled key used
in Definition~\ref{def:key-rate} implies that the key is both uniformly
distributed and independent of the adversary's knowledge, just as
in~\cite{RenKoe05}.  Previous works considered uniformity and security
separately. Note that, even though weaker than certain alternative
criteria such as the one of~\cite{DevWin04}, the security measure of
Definition~\ref{def:key-rate} is universally
composable~\cite{RenKoe05}.

In~\cite{Ben-OrHLMO05}, the question was posed whether the security
condition also holds if the accessible information is used instead of
the criterion considered here. Recently, it has been shown that this
is not the case~\cite{KRBM05}.  More precisely, an example of a family
of states was exhibited such that Eve has exponentially small
knowledge in terms of accessible information but constant knowledge in
terms of the Holevo information. This implies that in this context,
security definitions based on the accessible information are
problematic. In particular, a key might be insecure even though the
accessible information of an adversary on the key is exponentially
small (in the key size).


\section{Upper bounds for the key rate}\label{section-bounds}

In this section, we first derive sufficient conditions that a function
has to satisfy in order to be an upper bound for the key rate
(Section~\ref{sec:mono}). We focus on functions that are \emph{secrecy
  monotones}~\cite{CeMaSc02}, i.e., they are monotonically decreasing
under LOPC operations. Our approach therefore parallels the situation
in classical and quantum information theory where resource
transformations are also bounded by monotonic functions; examples
include the proofs of converses to coding theorems and entanglement
measures (see, e.g., \cite{Michal2001}). As a corollary to our
characterisation of secrecy monotones, we show how to turn
entanglement monotones into secrecy monotones.

In a second part (Section~\ref{sec:exmon}), we provide a number of
concrete secrecy monotones that satisfy the conditions mentioned
above. They can be roughly divided into two parts: (i) functions
derived from the intrinsic information and (ii) functions based on
entanglement monotones. Finally, we will compare different secrecy
monotones (Section~\ref{sec:compmon}) and study a few particular cases
in more detail (Section~\ref{sec:esep}).

\subsection{Secrecy monotones} \label{sec:mono}

\begin{theorem}\label{thm:secrecy-mon}
  Let $M(\rho)$ be a function mapping tripartite quantum states $\rho
  \equiv \rho_{ABE}$ into the positive numbers such that the following
  holds: \bee
\item Monotonicity: $M(\Lambda(\rho))\leq M(\rho)$ for any LOPC
  operation $\Lambda$.
\item Asymptotic continuity: for any states $\rho^n,\sigma^n$ on
  $\hcal_A^{n}\ot\hcal_B^{n}\ot\hcal_E^{n}$, the condition $\|\rho^n
  -\sigma^n\| \to 0$ implies $ {1\over \log r_n}\big|
  M(\rho^n)-M(\sigma^n)\big| \to 0 $ where $r_n=\dim (\hcal_A^{n} \ot
  \hcal_B^{n} \otimes \hcal_E^n)$.
\item Normalisation: $ M(\tau^\ell)=\ell \ .$ \eee Then the
  \emph{regularisation} of the function $M$ given by $
  M^\infty(\rho)=\limsup_{n \to \infty} {M(\rho^{\ot n})\over n} $ is
  an upper bound on $K_D$, i.e., $M^\infty(\rho_{ABE}) \geq
  K_D(\rho_{ABE})$ for all $\rho_{ABE}$ with $\dim \cH_A \otimes \cH_B
  \otimes \cH_E <\infty$.  If in addition $M$ satisfies \bee
\item[4.] Subadditivity on tensor products: $M(\rho^{\otimes n})\leq n
M(\rho)$, \eee then $M$ is an upper bound for $K_D$.
\end{theorem}

\begin{proof}
  Consider a key distillation protocol $\pcal$ that produces output
  states $\sigma^n$ such that $\|\sigma^n-\tau^{\ell_n}\| \rightarrow
  0$.  We will show that $M^\infty(\rho)\geq R_\pcal$. Let us assume
  without loss of generality that $R_\pcal>0$.  Indeed, by
  monotonicity we have $M(\rho^{\ot n})\geq M(\sigma^n)$, which is
  equivalent to \be \label{eq:limit} {1\over n} M(\rho^{\ot n})\geq
  {\ell_n \over n} \biggl( {M(\sigma^n) - M(\tau^{\ell_n})\over
    \ell_n} + 1 \biggr) \ ,\ee where we have used the normalisation
  condition.  As remarked in Definition~\ref{def:key-rate} there is a
  constant $c>0$ such that $\log r_n \leq c n$ and by definition of
  $R_\pcal$ there exists a $c'>0$ and $n_0$ such that for all $n\geq
  n_0$, $\log d_n \geq c' n$. Hence $\ell_n \geq c' n \geq
  \frac{c'}{c} \log r_n$, therefore asymptotic continuity implies
  \beopt \lim_{n \to \infty} {1\over \ell_n}\big|
  M(\sigma^n)-M(\tau^{\ell_n})\big| = 0 \ .\eeopt Taking the limsup
  on both sides of~(\ref{eq:limit}) gives $M^\infty(\rho)\geq
  \limsup_n {\ell_n \over n} = R_\pcal$. Thus we have shown that
  $M^\infty$ is an upper bound for the rate of an arbitrary protocol,
  so that it is an also upper bound for $K_D$.  
\end{proof}

If we restrict our attention to the special case of key distillation
from bipartite states $\rho_{AB}$, we can immediately identify a
well-known class of secrecy monotones, namely entanglement monotones.
A convenient formulation is in this case not given by the distillation
of states $\tau^{\ell}$ with help of LOPC operations, but rather by
the distillation of states $\gamma^\ell$ via local operations and
classical communication (LOCC), where $\gamma^\ell=U
\proj{\psi}_{AB}^{\otimes \ell} \otimes \rho_{A'B'} U^\dagger$, for
some unitary $U=\sum_{i=1}^{2^\ell} \proj{ii}_{AB} \otimes
U_{A'B'}^{(i)}$ and
$\ket{\psi}=\frac{1}{\sqrt{2}}(\ket{00}+\ket{11})$~\cite{pptkey,keyhuge}.
Note that measuring the state $\gamma^\ell$ with respect to the
computational bases on Alice and Bob's subsystems results in $\ell$
key bits.

\begin{corollary}\label{cor:entanglement-mon}
  Let $ E(\rho)$ be a function mapping bipartite quantum states $\rho\equiv \rho_{AB}$ into the positive numbers such that
  the following holds: \bee
\item Monotonicity: $E(\Lambda(\rho))\leq E(\rho)$ for any LOCC
  operation $\Lambda$.
\item Asymptotic continuity: for any states $\rho^n,\sigma^n$ on
  $\hcal_A^{n}\ot\hcal_B^{n}$, the condition $\|\rho^n -\sigma^n\| \to
  0$ implies ${1\over \log r_n}\big| E(\rho^n)-E(\sigma^n)\big| \to 0
  $ where $r_n=\dim (\hcal_A^{n} \ot \hcal_B^{n})$.
\item Normalisation: $ E(\gamma^\ell)\geq \ell \ .$ \eee Then the
  regularisation of the function $E$ given by $
  E^\infty(\rho)=\limsup_{n \to \infty} {E(\rho^{\ot n})\over n} $ is
  an upper bound on $K_D$, i.e., $E^\infty(\rho_{AB}) \geq
  K_D(\proj{\psi}_{ABE})$ where $\proj{\psi}_{ABE}$ is a purification
  of $\rho_{AB}$.  If in addition $E$ satisfies \bee
\item[4.] Subadditivity on tensor products: $E(\rho^{\otimes n})\leq n
  E(\rho)$, \eee then $E$ is an upper bound for $K_D$.

\end{corollary}

The analogue of this result in the realm of \emph{entanglement}
distillation has long been known: namely, every function $E$
satisfying LOCC monotonicity, asymptotic continuity near maximally
entangled states as well as normalisation on maximally entangled
states ($E(\proj{\psi})=\log d$ for $\ket{\psi}=\frac{1}{\sqrt{d}}
\sum_i \ket{ii}$) can be shown to provide an upper bound on
distillable entanglement $E_D$~\cite{limits,DonaldHR2001}, that is,
$E^\infty(\rho) \geq E_D(\rho)$. Additionally, if $E$ is subadditive,
the same inequality holds with $E^\infty$ replaced by $E$. Indeed this
result can be seen as a corollary to
Corollary~\ref{cor:entanglement-mon} by restricting from distillation
of states $\tau^\ell$ to distillation of $\proj{\psi}^{\otimes \ell}$
and noting that $\proj{\psi}^{\otimes \ell}$ is of the form
$\tau^\ell$ with trivial $A'B'$.

In the above corollary, we have identified asymptotic continuity on
\emph{all} states as well as normalisation on the states $\gamma^\ell$
(rather than on singlets) as the crucial ingredients in order for an
entanglement measure to bound distillable key from above.  Note also
that we require those additional conditions as, for instance, the
\emph{logarithmic negativity} as defined in~\cite{Vidal-Werner}
satisfies the weaker conditions, therefore being an upper bound on
distillable entanglement, but fails to be an upper bound on
distillable key.

We will now show how to turn this bound for bipartite states (or
tripartite pure states) into one for arbitrary tripartite states. The
recipe is simple: for a given state $\rho_{ABE}$, consider a
purification $\proj{\psi}_{AA'BB'E}$ where the purifying system is
denoted by $A'B'$ and is split between Alice and Bob. Clearly, for any
splitting, $K_D(\proj{\psi}_{AA'BB' E})\geq K_D(\rho_{ABE})$. This
inequality combined with the previous corollary applied to
$\proj{\psi}_{AA'BB' E}$ proves the following statement.
\begin{corollary} If $E$ satisfies the conditions of
  Corollary~\ref{cor:entanglement-mon} then \beopt K_D(\rho_{ABE})
  \leq E^\infty(\rho_{AA'BB'}) \ , \eeopt where
  $\rho_{AA'BB'}=\tr_{E} \proj{\psi}_{AA'BB'E}$ and $\rho_{ABE}=
  \tr_{A'B'} \proj{\psi}_{AA'BB'E}$.  If $E$ is subadditive, the same
  inequality holds with $E$ replacing $E^\infty$.
\end{corollary}

\subsection{Examples of secrecy monotones} \label{sec:exmon}

We will now introduce a number of secrecy monotones. We will only
briefly comment on the relations between them. A more detailed
analysis of how the different bounds on the key rate compare is given
in Section~\ref{sec:compmon}.

\subsubsection{Intrinsic information}

The \emph{intrinsic information} of a probability distribution
$p_{ijk}$ is given by \be\label{def-intr-original} I(A:B\downarrow
E):= \inf I(A:B|E')_\rho\ee where $\rho_{ABE}$ is the ccc state
corresponding to $p_{ijk}$. The infimum is taken over all channels
from $E$ to $E'$ specified by a conditional probability distributions
$p_{l|m}$. $\rho_{ABE'}$ is the state obtained by applying the channel
to $E$. This quantity has been defined by Maurer and Wolf and provides
an upper bound on the key rate from classical
correlations~\cite{MauWol97c-intr}. We can extend it in the following
way to arbitrary tripartite quantum states $\rho_{ABE}$.
\begin{definition} The \emph{intrinsic information} of a tripartite
quantum state $\rho_{ABE}$ is given by
 \beopt I(A:B\downarrow E)_\rho:=\inf I(A:B|E')_\rho \eeopt
 where the infimum is taken over all CPTP maps $\Lambda_{E\to E}$ from $E$ to
$E'$ where $\rho_{ABE'}= (I_{AB}\ot \Lambda_{E\to E} )(\rho_{ABE})$.
\end{definition}
This definition is compatible with the original definition since it
reduces to~(\ref{def-intr-original}) if the systems $A$, $B$ and $E$
are classical.

As shown in Appendix~\ref{app:intrprop}, the intrinsic information
satisfies the requirements of Theorem~\ref{thm:secrecy-mon} and,
hence, is an upper bound on the key rate.

\begin{theorem} \label{thm:intr-bound} The intrinsic
  information is an upper bound on distillable key, i.e.,
  $K_D(\rho_{ABE})\leq I(A:B\downarrow E)_\rho$.
\end{theorem}

Let us note that this bound differs from the bound proposed in
\cite{MoroderCL2005-povmintr,ChristandlR04-ABEkey} where instead of
all quantum channels, arbitrary measurements were considered.  Our
present bound can be tighter, as it can take into account Eve's
quantum memory.

In the case where $\rho_{ABE}$ is pure, this bound can be improved by
a factor of two because $I(A:B\downarrow E)_\rho =2E_{sq}(\rho_{AB})$,
where $E_{sq}$ is the squashed entanglement defined below and because
squashed entanglement is an upper bound for the key rate.

\subsubsection{Squashed entanglement}

\begin{definition}
 Squashed entanglement is defined as
\beopt E_{sq}(\rho_{AB})   = \frac{1}{2} \inf_{\substack{ \rho_{ABE}:\\
\rho_{AB}=\tr_E\rho_{ABE}}} I(A:B|E)_\rho \eeopt
\end{definition}
Squashed entanglement can be shown to be a LOCC monotone,
additive~\cite{ChrWin04}, and asymptotically
continuous~\cite{Alicki-Fannes}.
In~\cite[Proposition~4.19]{ChristandlPhD} it was shown to satisfy the
normalisation condition and is therefore an upper bound on distillable
key according to Corollary~\ref{cor:entanglement-mon}.

\begin{theorem} \label{thm:squashed} Squashed entanglement is an upper bound on distillable key, i.e., $K_D(\rho_{ABE})\leq E_{sq}(\rho_{AA'BB'})$ where
  $\rho_{AA'BB'}=\tr_{E} \proj{\psi}_{AA'BB'E}$ and $\rho_{ABE}=
  \tr_{A'B'} \proj{\psi}_{AA'BB'E}$.
\end{theorem}

\subsubsection{\Twointr} There is another way in which we can find a
bound on the key rate which is tighter than the intrinsic information.
In~\cite{RenWol03} it was shown that the classical intrinsic
information is \emph{E-lockable}, i.e., it can increase sharply when a
single bit is taken away from Eve. Since (classical) distillable key
is not E-lockable, the bound that the intrinsic information provides
cannot be tight. This was the motivation for defining the {\it
  \twointr} by $I(AB\downarrow\downarrow E)=\inf I(AB \downarrow E E')
+S(E')$ where the infimum is taken over arbitrary classical values
$E'$~\cite{RenWol03}.  We now define the quantum extension of this
function.

\begin{definition} Let $a=1,2$. The \emph{reduced intrinsic information (with parameter $a$)} is given by
  \beopt I(A:B\downarrow\downarrow E)^{(a)}_\rho=\inf \{ I(AB
  \downarrow E E')_\rho + a S(E')_\rho\} \eeopt where the infimum is
  taken over all extensions $\rho_{ABEE'}$ with a classical register
  $E'$ if $a=1$ and over arbitrary extensions $\rho_{ABEE'}$ if $a=2$.
\end{definition}

The parameter $a$ reflects the different behaviour of the intrinsic
information subject to loss of a single bit (qubit). The \twointr is
an upper bound on distillable key since \beopt K_D(\rho_{ABE})\leq
K_D(\rho_{ABEE'})+aS(E') \leq I(AB\downarrow EE') +aS(E') \ . \eeopt
The first inequality corresponds to Corollary~\ref{cor:e-lock} below.

\begin{theorem} \label{thm:intr} The reduced intrinsic information is an upper bound on distillable key, i.e., $K_D(\rho_{ABE})\leq I(A:B\downarrow\downarrow E)^{(a)}_\rho$,  for $a=1,2$.
\end{theorem}

\subsubsection{Relative entropy of entanglement}

The relative entropy of entanglement and its regularised version are
well-known entanglement measures that serve as important tools in
entanglement theory.

\begin{definition}
The relative entropy of entanglement is given
  by~\cite{VPRK1997,PlenioVedral1998} \beopt
  E_R(\rho_{AB})=\inf_{\sigma_{AB}}
  S(\rho_{AB}\|\sigma_{AB}) \eeopt where $S(\rho_{AB}\|\sigma_{AB})=\tr \rho_{AB} [\log \rho_{AB} -\log
  \sigma_{AB}]$ and the minimisation is taken over all separable states
  $\sigma_{AB}$, i.e. $\sigma_{AB}=\sum_i p_i \rho^i_A \otimes \rho^i_B$.
\end{definition}

The relative entropy of entanglement was the first upper bound that
has been provided for $K_D(\proj{\psi}_{ABE})$~\cite{pptkey,keyhuge}.
We now extend this result to all tripartite quantum states
$\rho^{ABE}$.

\begin{theorem} \label{thm:rel-bound} The relative entropy of
  entanglement is an upper bound on distillable key, i.e.,
  $K_D(\rho_{ABE})\leq E_R^\infty(\rho_{AA'BB'})\leq
  E_R(\rho_{AA'BB'})$ where $\rho_{AA'BB'}=\tr_{E}
  \proj{\psi}_{AA'BB'E}$ and $\rho_{ABE}= \tr_{A'B'}
  \proj{\psi}_{AA'BB'E}$.
\end{theorem}

It is a particular advantage of $E_R$ in its function as an upper
bound that it is not lockable~\cite{lock-ent}.

\subsection{Comparison of secrecy monotones} \label{sec:compmon}

\subsubsection{Pure versus mixed}

For entangled states, bounds derived from entanglement measures are
usually tighter than the intrinsic information and its reduced
version.  Consider for example the state $\rho_{ABE}=\proj{\psi}_{AB}
\ot \rho_E$ where $\ket{\psi}_{AB}={1\over \sqrt2}(|00\>+|11\>)$. Here
we have
\beopt E_R(\rho_{ABE})=E_R^\infty(\rho_{ABE})=E_{sq}(\rho_{ABE})=K_D(\rho_{ABE})=1\ , \eeopt
while
\beopt I(A:B\downarrow E)_\rho= I(A:B\downarrow\downarrow E)^{(a)}_\rho = 2
\ , \eeopt
for $a=1,2$.  In general, for tripartite pure states, squashed
entanglement is a tighter bound on the key rate than the intrinsic
information by at least a factor of two: \beopt
2E_{sq}(\proj{\psi}_{ABE})=I(A:B \downarrow E)_{\proj{\psi}} \ .  \eeopt

\subsubsection{The locking effect}

We will now give a concrete example which shows that there is a
purification $\ket{\psi}_{AA'BB'E}$ of $\rho_{ABE}$ such that
\beopt K_D(\rho_{ABE})=E_R(\rho_{AA'BB'})<I(AA':BB'\downarrow E)_\rho \ . \eeopt
Consider the distribution $p_{ijkl}$ defined by the following
distribution for $p_{ij}$ \vskip1mm \ben \nonumber
&&\begin{tabular}{|cr||c| c|c|c|}
  \hline & & & & & \\[-2ex]
  & $i$ \hspace{0.2em} &  $0$ & $1$ & $2$ & $3$\\
\hspace{1.2em} & \hspace{2em} & \hspace{2em} & \hspace{2em} & \hspace{2em} & \hspace{2em} \\[-1ex]  
$j$ &  &  &   &   &  \\ \hline \hline & & & & & \\[-2ex] $0$      & & ${1\over 8}$ &
${1\over 8}$ & $0$ & $0$ \\[-2ex] & & & & &  \\ \hline & & & & & \\[-2ex] $1$    &  & ${1\over 8}$ & ${1\over 8}$
& $0$ & $0$ \\[-2ex]
& & & & &  \\ 
\hline & & & & & \\[-2ex] $2$   &     &     $0$ & $0$ & ${1\over 4}$ & $0$\\[-2ex]  & & & & & \\ \hline & & & & & \\[-2ex] $3$ &  &
$0$ & $0$ & $0$ & ${1\over 4}$ \\[-2ex] & & & & & \\ \hline
\end{tabular}\een
and where $k$ and $l$ are uniquely determined by $(i,j)$, \ben
&&k =i + j (\text{mod } 2) {\quad \text{for} \quad} i,j\in\{0,1\} \\
&&k = i (\text{mod } 2) {\quad \quad \text{for} \quad} i \in \{2,3\} \\
&&l = \lfloor i/2 \rfloor \een for all $(i,j)$ with $p_{ij}>0$. We
denote the corresponding cccc state by \beopt \rho_{ABEF}=\sum_{ijkl}
p_{ijkl} \proj{ijkl} \ .  \eeopt Clearly $K_D(\rho_{A;B;EF})=0$, as
Eve can factorise Alice and Bob, by keeping $k$ when $l=1$ and
forgetting it when $l=0$. In the former case, when $l=0$, then Alice
and Bob have $(i,j)=(2,2)$, and when $l=1$, then Alice and Bob have
$(i,j)=(3,3)$. In the latter case, both Alice and Bob have at random
$0$ or $1$ and they are not correlated.

On the other hand, when Eve does not have access to $l$, then the key
rate is equal to $1$, i.e., $K_D(\rho_{A;B;E})$=1. Indeed, it cannot
be greater, as key cannot increase more than the entropy of the
variable that was taken out from Eve. However one finds that the
intrinsic information is equal to $3/2$, i.e., ${I(A:B\downarrow
  E)_\rho}=3/2$~\cite{RenWol03}.

Let us consider the purification of the above state, \ben |\psi_{A'A B
  EF}\>&=&{1\over 2} \bigl(|0\>_{A'} |22\>_{AB} |0\>_E|0\>_F
+ |0\>_{A'} |33\>_{AB} |1\>_E|0\>_F   \\
&& \quad +|\psi\>_{A'AB} |0\>_E |1\>_F + |\phi\>_{A'AB} |1\>_E |1\>_F
\bigr),\nonumber \een where \beopt
|\psi\>={1\over \sqrt2} (|0\>_{A'}
|00\>_{AB} + |1\>_{A'}|11\>_{AB}) \eeopt
and \beopt |\phi\>={1\over \sqrt2}
(|0\>_{A'} |01\>_{AB} + |1\>_{A'}|10\>_{AB})\ . \eeopt
Thus when $E$ and
$F$ are with Eve, the state $\rho_{AA';B}$ of Alice and Bob is a
mixture of four states: $|0\>|22\>$, $|0\>|33\>$, $|\phi\>$ and $|\psi
\>$.  This state is separable state, hence $E_R(\rho_{AA';B})=0$.

Consider now the state $\rho_{AA'F;B}$ where $F$ is controlled by
Alice instead of Eve. Measuring $F$ makes the state separable and in
\cite{lock-ent} it was shown that measuring a single qubit cannot
decrease the relative entropy of entanglement by more than $1$, thus
we obtain \beopt E_R(\rho_{AA'F;B})\leq 1 \ . \eeopt By
Theorem~\ref{thm:rel-bound} we then have $K_D(\rho_{ABE})\leq 1$, but
indeed one can distil one bit of key from $\rho_{ABE}$, therefore
\beopt K_D(\rho_{ABE})=E_R(\rho_{AA'F;B})=1 \ . \eeopt
In~\cite{RenWol03} the considered distribution was generalised to make
the gap between intrinsic information and distillable key arbitrarily
large. It is not difficult to see that $E_R$ is still bounded by one.
This shows that the bound based on relative entropy of entanglement,
though perhaps more complicated in use, can be significantly stronger
than intrinsic information bound. We leave it open, whether or not the
intrinsic information bound is weaker in general when compared to the
relative entropy bound. This parallels the challenge to discover a
relation between the relative entropy of entanglement and squashed
entanglement. Here it has also been observed that squashed
entanglement can exceed the relative entropy of entanglement by a
large amount, due to a \emph{locking effect}~\cite{ChristandlW-lock}.

\subsection{Upper and lower bounds when $\rho_{ABE}=\rho_{AB}\otimes \rho_E$}
\label{sec:esep}

In this section we focus on states of the form
$\rho_{ABE}=\rho_{AB}\ot \rho_E$.  Since distillable key cannot
increase under Eve's operations, the form of the state $\rho_E$ is not
important and we conclude that $\kd(\rho_{AB}\otimes \rho_E)$ is a
function of $\rho_{AB}$ only.  If the state $\rho_{AB}$ is classical
on system $A$, then it is known that distillable key is equal to the
quantum mutual information, $\kd(\rho_{AB}\otimes \rho_E) =
I(A:B)_\rho$ \cite{DevWin04}. Indeed, we know from
Theorem~\ref{thm:intr-bound} that the key rate can never exceed
$I(A:B)_\rho$. For separable quantum states $\rho_{AB}$ we were able
to further improve this bound. The upper bounds are summarised in the
following theorem, whose proof is given in Appendix~\ref{app:boundpr}.

\begin{theorem} \label{prop-upper-separable}
For all states $\rho_{AB}\otimes \rho_E$,
\beopt \kd(\rho_{AB}\otimes \rho_E) \leq I(A:B)_\rho \eeopt
with equality if
$\rho_{AB}$ is classical on system $A$. If $\rho_{AB}$ is separable,
i.e., $\rho_{AB}=\sum_i p_i \rho^i_A \otimes \rho^i_B$, then 
\begin{eqnarray} 
\kd(\rho_{AB} \otimes \rho_E) &\leq & I_{\acc}^{\LOPC}(\cE)\leq
I_{\acc} (\cE) 
\end{eqnarray}
where $\cE=\{p_i, \rho_{A}^{i}\ot \rho_{B}^{i}\}$ and
$I_{\acc}^{\LOPC}(\cE)$ is the maximal mutual information that Alice
and Bob can obtain about $i$ using LOPC operations (see e.g.
\cite{Bennett-nlwe,Badziag-Holevo}), whereas $I_{\acc}(\cE)$ denotes
the usual accessible information, i.e. maximal mutual information
about $i$ obtained by joint measurements.
\end{theorem}

We will now derive a general lower bound on the key rate in terms of
the distillable common randomness.

\begin{definition}
  We say that an LOPC protocol $\cP$ distills \emph{common randomness}
  at rate $\cR_\cP$ if there exists a sequence $\{\ell_n\}_{n \in
    \mathbb{N}}$ such that
  \begin{align} 
  \limsup_{n \to \infty} {\ell_n-m_n\over  n} & = R_\pcal \\
  \lim_{n\to \infty} \|\Lambda_n(\rho^{\ot n}_{AB}) - \tau^{\ell_n}\| & = 0 
  \end{align}
  where $m_n$ is the number of communicated bits. The \emph{distillable
    common randomness} of a state $\rho^{AB}$ is defined as
  $D_R(\rho_{AB}) :=\sup_{\cP} \cR_\cP$.
\end{definition}

For some protocols the rate may be negative. However it is immediate
that $\dr(\rho_{AB})$ is nonnegative for all $\rho_{AB}$. The
following statement is a direct consequence of the results
in~\cite{DevWin04,RenKoe05}.
\begin{theorem}
  For the states $\rho_{ABE}=\rho_{AB}\ot \rho_E$ the distillable key
  is an upper bound on the distillable common randomness, i.e.,
  $K_D(\rho_{AB}\ot \rho_E)\geq \dr(\rho_{AB})$ for all $\rho_{AB}$
  and $\rho_E$.
\end{theorem}

\section{Embedding classical into quantum states} \label{section-embedding}

The problem of distilling key from a classical tripartite distribution
(i.e., ccc states) is closely related to the problem of distilling
entanglement from a bipartite quantum state (where the environment
takes the role of the adversary), as noted
in~\cite{GisWol00,RenWol03}. It thus seems natural to ask whether, in
analogy to \emph{bound entangled} quantum states (which have positive
entanglement cost but zero distillable entanglement), there might be
classical distributions with \emph{bound information}.  These are
distributions with zero key rate but positive key cost, i.e., no key
can be distilled from them, yet key is needed to generate them.  The
existence of such distributions, however, is still unproved.  (There
are, however, some partial positive answers, including an asymptotic
result~\cite{RenWol03} as well as a result for scenarios involving
more than three parties~\cite{AcinCM-boundinfo}.)

In~\cite{GisWol00,RenWol03}, it has been suggested that the classical
distribution obtained by measuring bound entangled quantum states
might have bound information. Such hope, however, was put into
question by the results of~\cite{pptkey}, showing that there are
quantum states with positive key rate but no distillable entanglement
(i.e., they are bound entangled). However, the examples of states put
forward in~\cite{pptkey} have a rather special structure. It is thus
still possible that distributions with bound information might be
obtained by measuring appropriately chosen bound entangled states.


In the following, we consider a special \emph{embedding} of classical
distributions into quantum states as proposed in~\cite{GisWol00}. We
then show how statements about key distillation starting from the
original state and from the embedded state are related to each other.
Let
\begin{equation} \label{eq:ccc}
  \rho_{ccc} := \sum_{ijk} p_{ijk} \proj{ijk}_{ABE}
\end{equation}
be a ccc state defined relative to fixed orthonormal bases on the
three subsystems (in the following called \emph{computational bases}).
We then consider the \emph{qqq embedding} $\rho_{qqq} = \proj{\psi}$
of $\rho_{ccc}$ given by
\beopt
\ket{\psi}=\sum_i \sqrt{p_{ijk}} \ket{ijk}_{ABE} \ .
\eeopt
Note that, if Alice and Bob measure $\rho_{qqq}$ in the computational
basis, they end up with a state of the form
\begin{equation} \label{eq:embccq}
  \rho_{ccq}
=
  \sum_{ij}p_{ij} \proj{ij}_{AB}\otimes \proj{\psi^{ij}}_E
\end{equation}
for some appropriately chosen $\ket{\psi^{ij}}$. We call this state
the \emph{ccq embedding} of $\rho_{ccc}$.

In a similar way as classical distributions can be translated to
quantum states, classical protocols have a quantum analogue. To make
this more precise, we consider a classical LOPC protocol $\cP$ that
Alice and Bob wish to apply to a ccc state $\rho_{ccc}$ as
in~\eqref{eq:ccc}.  Obviously, $\cP$ can equivalently be applied to
the corresponding ccq embedding $\rho_{ccq}$ as defined
in~\eqref{eq:embccq} (because Alice and Bob's parts are the same in
both cases). Because Eve might transform the information she has in
the ccq case to the information she has in the ccc case by applying a
local measurement, security of the key generated by $\cP$ when applied
to $\rho_{ccq}$ immediately implies security of the key generated by
$\cP$ when applied to $\rho_{ccc}$.  Note, however, that the opposite
of this statement is generally not true.

In general, a classical protocol $\cP$ can be subdivided into a
sequence of steps of the following form:
\begin{enumerate}
\item generating local randomness
\item forgetting information (discarding local subsystems)
\item applying permutations
\item classical communication.
\end{enumerate}
The \emph{coherent version} of $\cP$, denoted $\cP_{q}$, is defined as
the protocol acting on a qqq state where the above classical
operations are replaced by the following quantum operations:
\begin{enumerate}
\item attaching subsystems which are in a superposition of fixed basis
  vectors
\item transferring subsystems to Eve
\item applying unitary transformations that permute fixed basis
  vectors
\item adding ancilla systems (with fixed initial state) to both the
  receiver's and Eve's system, and applying controlled not (CNOT)
  operations to both ancillas, where the CNOTs are controlled by the
  communication bits.
\end{enumerate}

Consider now a fixed ccc state $\rho_{ccc}$ of the form~\eqref{eq:ccc}
and let $\cP$ be a classical protocol acting on $\rho_{ccc}$.  It is
easy to see that the following operations applied to the qqq embedding
$\rho_{qqq}$ of $\rho_{ccc}$ result in the same state: (i)~measuring
in the computational basis and then applying the classical protocol
$\cP$; or (ii)~applying the coherent protocol $\cP_q$ and then
measuring the resulting state $\gamma^\ell$ in the computational basis.
This fact can be expressed by a commutative diagram.
\beopt
    \begin{CD}
      \proj{\psi}^{\otimes n} @>\cP_{q}>>  \gamma^\ell \\
      @V\mbox{measurement}VV @VV\mbox{measurement}V \\
      \rho_{ccq}^{\otimes n} @>\cP>> \tau^{\ell}
    \end{CD}
\eeopt
Hence, if the coherent version $\cP_q$ of $\cP$ acting on $\rho_{qqq}$
distills secure key bits at rate $R$ then so does the protocol $\cP$
applied to the original ccc state $\rho_{ccc}$.

It is natural to ask whether there are cases for which the converse of
this statement holds as well. This would mean that security of a
classical protocol also implies security of its coherent version. In
the following, we exhibit a class of distributions for which this is
always true. The key rate of any such distribution is thus equal to
the key rate of the corresponding embedded qqq state.

Roughly speaking, the class of distributions we consider is
characterised by the property that the information known to Eve is
completely determined by the joint information held by Alice and Bob.

\begin{theorem} \label{thm:classq}
  Let $\rho_{ccc}$ be a ccc state of the form~\eqref{eq:ccc} such
  that, for any pair of values $(i,j)$ held by Alice and Bob there
  exists at most one value $k$ of Eve with $p_{ijk} > 0$. If a
  classical protocol $\cP$ applied to $\rho_{ccc}$ produces key at
  rate $R$ then so does its coherent version $\cP_q$ applied to the
  qqq embedding $\ket{\psi}$ of $\rho_{ccc}$ (and followed by a
  measurement in the computational basis).
\end{theorem}

\begin{proof}
  The ccq embedding of $\rho_{ccc}$ is given by a state of the form
\beopt
  \rho_{ccq}
=
  \sum_{ij}p_{ij} \proj{ij}_{AB}\otimes \proj{\psi^{ij}}_E \ .
\eeopt
Since, by assumption, every pair $(i,j)$ determines a unique $k =
k(i,j)$, $\proj{\psi^{ij}}_E$ equals $\proj{k(i,j)}$ and, hence,
$\rho_{ccq}$ is identical to the original ccc state $\rho_{ccc}$. The
assertion then follows from the fact that measurements in the
computational basis applied to Alice and Bob's subsystems commute with
the coherent version $\cP_q$ of $\cP$. 
\end{proof}

\begin{corollary}
  Let $\rho_{ccc}$ be a ccc state of the form~\eqref{eq:ccc} such
  that, for any pair of values $(i,j)$ held by Alice and Bob there
  exists at most one value $k$ of Eve with $p_{ijk} > 0$. Then, the
  key rate for the qqq embedding $\rho_{qqq}$ of $\rho_{ccc}$
  satisfies
  \beopt
    K_D(\rho_{qqq}) = K_D(\rho_{ccc}) \ .
  \eeopt
\end{corollary}

Note that the above statements do not necessarily hold for general
distributions.  To see this, consider the state
\beopt
  \ket{\psi}_{ABA'E}
=
  \ket{00}_{AB} \ket{+}_{A'}\ket{+}_{E} + \ket{11}_{AB} \ket{\psi_+}_{A'E}
\eeopt
where $\ket{+}:={1\over \sqrt2} (\ket{0}+\ket{1})$ and
$\ket{\psi_+}:={1\over \sqrt2} (\ket{0}\ket{0}+\ket{1}\ket{1})$.
Moreover, let $\rho_{ccc}$ be the ccc state obtained by measuring
$\proj{\psi}_{AA';B;E}$ in the computational basis. Because all its
coefficient are positive, it is easy to verify that
$\proj{\psi}_{ABA'E}$ can be seen as the qqq embedding of
$\rho_{ccc}$. Observe that, after discarding subsystem $A'$,
$\rho_{ccc}$ corresponds to a perfect key bit. However, the ccq state
obtained from $\proj{\psi}_{ABA'E}$ by discarding $A'$ and measuring
in the computational basis is of the form ${1\over 2} (\proj{00}_{AB}
\ot \proj{+}_E + \proj{11}_{AB} \ot \id_E/2)$. This state, of course,
does not correspond to a key bit as Eve might easily distinguish the
states $\proj{+}$ and $\id_E/2$.

We continue with a statement on the relation between the intrinsic
information of a ccc state and the so-called \emph{entanglement of
  formation}\footnote{The \emph{entanglement of formation} $E_F$ is an
  entanglement measure defined for bipartite states by
  $E_F(\sigma_{AB}):=\min \sum_i p_i S(\tr_B(\sigma_{AB}^i))$ where
  the minimum is taken over all ensembles $\{p_i, \sigma_{AB}^i\}$
  with $\sum_i p_i \sigma_{AB}^i=\sigma_{AB}$~\cite{BDSW1996}.}  $E_F$
of its qqq embedding.  More precisely, we show that, under the same
condition as in Theorem~\ref{thm:classq}, the first is a lower bound
for the latter (see also~\cite{Christandl:2002,Renner00}).

\begin{theorem} \label{pr:emz}
  Let $\rho_{ccc}$ be a ccc state of the form~\eqref{eq:ccc} such
  that, for any pair of values $(i,j)$ held by Alice and Bob there
  exists at most one value $k$ of Eve with $p_{ijk} > 0$, and let
  $\rho_{qqq}$ be the qqq embedding of this state. Then
  \beopt
    I(A:B\downarrow E)_{\rho_{ccc}} \leq E_F(\tr_E(\rho_{qqq})) \ .
  \eeopt
\end{theorem}

\begin{proof}
  Note first that any decomposition of $\tr_E(\rho_{qqq})$ into pure
  states can be induced by an appropriate measurement on the system
  $E$. Hence, we have
  \begin{equation} \label{eq:Eof}
    E_F(\tr_E(\rho_{qqq}))
  =
    \min_{\{\ket{\bar{k}\}}} \sum_{\bar{k}} p_{\bar{k}} S(A)_{\ket{\psi_{\bar{k}}}}
  \end{equation}
  where the minimum ranges over all families of (not necessarily
  normalised) vectors $\ket{\bar{k}}$ such that $\sum_{\bar{k}}
  \proj{\bar{k}} = \id_E$ (this ensures that they form a measurement),
  $p_{\bar{k}} :=|\bra{\bar{k}}_E \ket{\psi}_{ABE}|^2$, and
  $\ket{\psi_{\bar{k}}}:=\bra{\bar{k}}_E\ket{\psi}_{ABE}/\sqrt{p_{\bar{k}}}$.

  For any pair $(i,j)$ of values held by Alice and Bob (with nonzero
  probability) we have $ \tr_{AB}\left[ \rho_{qqq} \left(\proj{ij}
      \otimes \id_E\right)\right]=p_{ij}\proj{k}$, where $k = k(i,j)$
  is the corresponding (unique) value held by Eve.  Hence, the
  probability distribution of the state $\bar{\rho}_{ccc}$ obtained by
  applying the above measurement on Eve's system satisfies
  \beopt
      q_{ij\bar{k}}
            :=\tr( \proj{\psi}_{ABE}  \proj{ij\bar{k}})
            =p_{ijk} q_{\bar{k}|k} \ ,
   \eeopt
   where $q_{\bar{k}|k} :=\tr(\proj{\bar{k}}\proj{k})$. The intrinsic
   information is thus bounded by
   \beopt
   I(A:B\downarrow E)_{\rho_{ccc}}
   \leq
   \min_{\{ \ket{\bar{k}}\}} I(A:B|\bar{E})_{\bar{\rho}_{ccc}} \ ,
  \eeopt
  where $\bar{\rho}_{ccc}$ is the state defined above (depending on
  the choice of the vectors $\ket{\bar{k}}$).  Moreover, using
  Holevo's bound, we find
  \beopt
    I(A:B|\bar{E})_{\bar{\rho}_{ccc}}
   \leq  \min_{\{\ket{\bar{k}\}}} \sum_{\bar{k}} p_{\bar{k}} S(A)_{\ket{\psi_{\bar{k}}}} \ .
  \eeopt
  The assertion then follows from~\eqref{eq:Eof}. 
\end{proof}

Because the intrinsic information is additive (i.e., it is equal to
its regularised version), Theorem~\ref{pr:emz} also holds if the
entanglement of formation $E_F$ is replaced by the entanglement cost
$E_C$.

The discussion above suggests that classical key distillation from ccc
states can indeed by analysed by considering the corresponding qqq
embedding of the state, but the original ccc state has to satisfy
certain properties. This relation might be particularly useful for the
study of bound information as discussed at the beginning of this
section. In fact, there exist bound entangled states which satisfy the
property required by Theorem~\ref{thm:classq}
above~\cite{PawelMaciek2000}.

\section{On locking and pre-shared keys} \label{sec:lock}

In \cite{RenWol03} it was observed that, by adding one bit of
information to Eve, the (classical) intrinsic information can decrease
by an arbitrarily large amount. In \cite{DiVincenzo-locking} it was
shown that classical correlation measures of quantum states can
exhibit a similar behaviour; more precisely, the accessible
information can drop by an arbitrarily large amount when a single bit
of information is lost. This phenomenon has been named \emph{locking
  of information} or just \emph{locking}. For tripartite states
$\rho_{ABE}$, locking comes in two flavours: i) locking caused by
removing information from Eve, ii) locking caused by removing
information from Alice and/or Bob (and possibly giving it to Eve). Let
us call those variants \emph{E-locking} and \emph{AB-locking},
respectively.

In \cite{lock-ent} it was shown that entanglement cost as well as many
other entanglement measures can be AB-locked.  Further results show
that squashed entanglement and entanglement of purification are also
AB-lockable \cite{ChristandlW-lock,Winter-keycost}.  So far the only
known non-lockable entanglement measure is relative entropy of
entanglement.

It was shown in~\cite{RenWol03} that distillable key is not E-lockable
for classical states.  In the sequel we extend this result and prove
that the distillable key for quantum states $\rho_{ABE}$ is not
E-lockable, either. The proof proceeds along the lines of
\cite{RenWol03}, replacing the bound of Csisz\'ar and K\"orner by its
quantum generalisations due to \cite{DevWin04} (see also
\cite{RenKoe05}).
Let us emphasise that we leave open the question on whether
distillable key is AB-lockable (even for ccc states).


\begin{theorem}
\label{thm:e-lock} Consider a state $\rho_{ABEE'}$ and let $\pcal$ be a key distillation
protocol for $\rho_{ABE}$ with rate $R_\pcal$. Then there exists
another protocol $\pcal'$ for $\rho_{ABE E'}$ with rate $R_{\pcal'}
\geq R_\pcal - 2 S(\rho_{E'})$. If, in addition, $E'$ is classical
then $R_{\pcal'} \geq R_\pcal - S(\rho_{E'})$.
\end{theorem}

\begin{proof}
  For any fixed $\ep > 0$ there exists $n \in \mathbb{N}$ such that
  the protocol $\pcal$ transforms $\rho_{ABE}^{\ot n}$ into a ccq
  state $\sigma_{ABE}$ which satisfies the following inequalities: \be
  \|\sigma_{ABE}-\tau^\ell\|\leq \ep, \quad {\ell\over n}\geq
  R_\pcal-\ep. \label{eq:almost-key} \ee Suppose that Alice and Bob
  apply this map to the state $\rho_{ABEE'}^{\ot n}$ (i.e., they try
  to distil key, as if the system $E'$ was not present).
  The state $\rho_{ABEE'}^{\ot n}$ is then transformed into some state
  $\sigma_{ABEE'}$ which traced out over $E'$ is equal to the ccq
  state $\sigma_{ABE}$.  Repeating this protocol $m$ times results in
  $\sigma_{ABEE'}^{\otimes m}$, from which Alice and Bob can draw at
  least $m(I(A:B)- I(A:EE')) - o(m)$ bits of key by error correction
  and privacy amplification~\cite{DevWin04}.  This defines a protocol
  $\pcal'$.  To evaluate its rate, we use subadditivity of entropy
  which gives the estimate \beopt
  I(A:EE')_\sigma\leq I(A:E)_\sigma +
  I(AE:E')_\sigma \ .
  \label{eq:est} \eeopt
  From~\eqref{eq:almost-key} and the conditional version of Fannes'
  inequality~\cite{Alicki-Fannes} we know that, for any $\ep \in
  [0,1]$,\footnote{$H(\ep)$ denotes the binary entropy, i.e., the
    Shannon entropy of the distribution~${[\ep, 1-\ep]}$.}
  \beopt I(A:B)_\sigma  -  I(A:E)_\sigma \geq (1- 8 \ep) \ell - 4 H(\ep) \ .
    \eeopt
    This together with \eqref{eq:lowerbound} implies \beopt
    K_D(\sigma_{ABEE'}) \geq I(A:B)_{\sigma} - I(A:EE')_{\sigma} \geq
    (1-8 \ep) \ell - 4 H(\ep) - I(AE:E')_{\sigma} \ .  \eeopt
    To get the
    key rate of $\pcal'$, we divide the above by $n$ and
    use~\eqref{eq:almost-key},
    \beopt R_{\pcal'} \geq {1\over n} K_D(\sigma_{ABEE'}) \geq (1- 8
    \ep)(R_{\pcal}-\ep) - {1\over n} 4 H(\ep) - {1\over n}
    I(AE:E')_{\sigma} \ .\eeopt Because this holds for any $\ep > 0$, the
    assertion follows from $ {I(AE:E')_\sigma} \leq 2S(E')_\sigma = 2
    n S(E')_\rho $ and, if $E'$ is classical, $ I(AE:E')_\sigma\leq
    S(E')_\sigma = n S(E')_\rho$.  
\end{proof}

Applying the above theorem to an optimal protocol leads to the
statement that the key rate $K_D$ is not E-lockable.

\begin{corollary} \label{cor:e-lock}
  For any state $\rho_{ABEE'}$, $K_D(\rho_{ABEE'}) \geq
  K_D(\rho_{ABE}) - 2S(\rho_{E'})$ and, if $E'$ is classical,
  $K_D(\rho_{ABEE'}) \geq K_D(\rho_{ABE}) - S(\rho_E')$.
\end{corollary}

Consider now a situation where Alice and Bob have some pre-shared key
$U$ which is not known to Eve.

A major consequence of Theorem~\ref{thm:e-lock} is that a pre-shared
key cannot be used as a catalyst to increase the key rate. More
precisely, the corollary below implies that, for any protocol $\pcal$
that uses a pre-shared key held by Alice and Bob, there is another
protocol $\pcal'$ which is as efficient as $\pcal'$ (with respect to
the net key rate), but does not need a pre-shared key.

\begin{corollary} \label{cor:lock}
  Let $\pcal$ be a key distillation protocol for $\rho_{ABE} \otimes
  \tau^\ell$ where $\tau^\ell$ is some additional $\ell$-bit key
  shared by Alice and Bob. Then there exists another protocol $\pcal'$
  for $\rho_{ABE}$ with rate $R_{\pcal'} \geq R_\pcal - \ell$.
\end{corollary}

\begin{proof}
  Consider the state $\rho_{A' B' E E'}$ where $E'$ is a system
  containing the value $U$ of a uniformly distributed $\ell$-bit key,
  $A' := (A,U)$, and $B' := (B,U)$. Note that $\rho_{A' B' E}$ is
  equivalent to $\rho_{ABE} \otimes \tau^\ell$. The assertion then
  follows from the observation that any protocol which produces a
  secure key starting from $\rho_{A' B' E E'}$ can easily be
  transformed into an (equally efficient) protocol which starts from
  $\rho_{A B E}$, because Alice and Bob can always generate public
  shared randomness. 
\end{proof}

The following example shows that the factor $2$ in
Theorem~\ref{thm:e-lock} and Corollary~\ref{cor:e-lock} is strictly
necessary. Let \beopt \rho_{ABEE'}=\sum_{i=1}^4 |i\rangle \langle i|_A
\otimes |i\rangle \langle i|_B \otimes |\psi_i \rangle \langle
\psi_i|_{EE'}  \eeopt where $\ket{\psi_i}$ are the four Bell states on the
bipartite system $EE'$.  Then, obviously, ${K_D(\rho_{A B E E'})}=0$,
but if $E'$ (which is only {\it one} qubit) is lost, then
${K_D(\rho_{A B E})}=2$, since $E$ is then maximally mixed conditioned
on $i$.  One recognises here the effect of superdense coding.

\section{Classical and quantum adversaries in {QKD}}
\label{section-qkd}




Up to now, we have considered an adversary with unbounded resources.
Of course, if one limits the adversary's capabilities, certain
cryptographic tasks might become easier. In the following, we will
examine a situation where the adversary cannot store quantum states
and, hence, is forced to apply a measurement, turning them into
classical data. We will exhibit an example of a $2 d$-dimensional
ccq state which only has key rate $1$, but if Eve is forced to
measure her system, the key rate raises up to roughly ${1\over 2}
\log d$.

Note that upper bounds on the key rate which are defined in terms of
an optimal measurement on Eve's system (see, e.g.,
\cite{MoroderCL2005-povmintr,ChristandlR04-ABEkey} and
Section~\ref{section-bounds}) are also upper bounds on the key rate in
a setting where Eve has no quantum memory.  Hence, our result implies
that these upper bounds are generally only rough estimates for the key
rate in the unbounded scenario.

Consider the state
\begin{multline} 
  \rho_{A A' B B' E} = {1\over 2d}\sum_{k=1}^d \proj{00}_{AB} (\proj{kk}_{A'B'}\ot
  \proj{k}_{E}) \\ + \proj{11}_{AB} (\proj{kk}_{A'B'}\ot
  U\proj{k}_E|U^\dagger)
\end{multline}
where $U$ is the quantum Fourier transform on $d$ dimensions. (Such a
state has been proposed in~\cite{DiVincenzo-locking} to exhibit a
locking effect of the accessible information. It also corresponds to
the \emph{flower state} of~\cite{lock-ent}.)

It is easy to see that the bit in the system $A B$ is uncorrelated to
Eve's information and, hence, completely secret, i.e., $K_D(\rho_{A A'
  B B' E}) = K_D(\rho_{A B}) \geq 1$. On the other hand, if this bit
is known to Eve then she has full knowledge on the state in $A' B'$,
i.e., $K_D(\rho_{A A' B B' E E'}) \leq I(A A':B B'\downarrow E
E')_\rho = 0$, where $E'$ is a classical system carrying the value of
the bit in $A B$ (see Theorem~\ref{thm:intr-bound}). From this and
Corollary~\ref{cor:e-lock} (or, alternatively,
Theorem~\ref{thm:intr}), we conclude that the key rate (relative to an
unbounded adversary) is given by
\beopt
  K_D(\rho_{A A' B B' E}) = K_D(\rho_{A B}) = 1 \ .
\eeopt

Let us now assume that Eve applies a measurement on her system $E$,
transforming the state defined above into a ccc state $\sigma_{A A' B
  B' E}$. Because the values of Alice and Bob are maximally
correlated, it is easy to see that the key rate of this state
satisfies $K_D(\sigma_{A A' B B' E}) = S(A|E)_{\sigma} = S(A)_{\sigma}
- I(A:E)_{\sigma}$.  Note that $S(A)_{\sigma} = 1 + \log d$. Moreover,
the mutual information $I(A:E)_{\sigma}$ for an optimal measurement on
$E$ corresponds to the so-called accessible information, which equals
${1\over 2}\log d$, as shown in~\cite{DiVincenzo-locking}. We thus
conclude that
\beopt
K_D(\sigma_{A A' B B' E}) = 1+ {1\over 2} \log d \ .
\eeopt
Note that the accessible information is additive, so even if the
measurements are applied to blocks of states, the amount of key that
can be generated is given by this expression.


The above result gives some insights into the strength of attacks
considered in the context of quantum key distribution (QKD).  A
so-called \emph{individual attack} corresponds to a situation where
the adversary transforms his information into classical values. In
contrast, a \emph{collective attack} is more general and allows the
storage of quantum states.

As shown in~\cite{Renner05}, for most QKD protocols, security against
collective attacks implies security against any attack allowed by the
laws of quantum physics. The above result implies that the same is not
true for individual attacks, i.e., these might be arbitrarily weaker
than collective (and, hence, also general) attacks.

%
%
%
%

\section*{Acknowledgment}

We are grateful to Karol Horodecki and Norbert L\"utkenhaus for their
valuable input and many enlightening discussions.  We would also like
to thank Hoi-Kwong Lo and anonymous reviewers for their helpful
comments and suggestions. Ppart of this work was completed during the
Isaac Newton Institute QIS programme 2004. The work was supported by
the European Commission through the FP6-FET Integrated Projects SCALA
CT-015714 and QAP IST-3-015848, QIP IRC (GR/S821176/01), and through
SECOQC.  MC acknowledges the support of an EPSRC Postdoctoral
Fellowship and a Nevile Research Fellowship, which he holds at
Magdalene College Cambridge. MH is supported by the Polish Ministry of
Scientific Research and Information Technology under grant no.\ 
PBZ-MIN-008/P03/2003. RR is supported by HP Labs Bristol.

\appendix

\section{On the definition of the key rate} \label{app:commlin}

The following lemma achieves a simplification of the definition of
distillable key~$K_D$ (Definition~\ref{def:key-rate}).

\begin{lemma}\label{lemma:linear}
  The maximisation in the definition of $K_D$ can be restricted to
  protocols that use communication at most linear in the number of
  copies of $\rho^{ABE}$.
\end{lemma}
\begin{proof}
Let $\{\Lambda_n\}_{n \in \mathbb{N}}$ be a key distillation
protocol with rate $R$ and with communication not necessarily linear
in $n$. Fix $\epsilon >0$. Then there exists an $n_0$ such that
\beopt || \Lambda_{n_0}(\rho_{ABE}^{\otimes n_0}) - \tau^{\ell_{n_0}}_{ABE} ||_1 \leq
\epsilon \eeopt and  $\frac{\ell_{n_0}}{n_0} \geq R-\epsilon$. Consider
now key distillation from many copies of $\sigma_{ABE}:=\Gamma
(\Lambda_{n_0}(\rho_{ABE}^{\otimes n_0}))$, where $\Gamma$ is a
measurement of Alice and Bob in their computational bases. We can
limit the dimension of Alice's and Bob's system to $2^{\ell_{n_0}}$,
because any additionally appearing symbols could be mapped, for
instance, to the symbol 1. This allows us to bound the difference in
the mutual informations with help of a conditional version of
Fannes' inequality~\cite{Alicki-Fannes},
\beopt
I(A:B)_\sigma -I(A:E)_\sigma \geq (1 - 8 \epsilon )\ell_{n_0} -4
H(\epsilon) \  , \eeopt
which holds if $\ep \leq 1$.  Alice and Bob can achieve
the rate $I(A:B)_\sigma -I(A:E)_\sigma$ using communication linear in
the number of copies of $\sigma_{ABE}$, since $\sigma_{ABE}$ is
evidently a ccq state~\cite{DevWin04,RenKoe05}. We have therefore
modified the protocol $\{\Lambda_n\}_{n \in \mathbb{N}}$ achieving a
rate $R$ into a protocol $\{\tilde\Lambda_n\}_{n \in \mathbb{N}}$ with
a rate
\beopt \tilde R\geq (1-8 \epsilon) (R-\epsilon) - \frac{4 H(\epsilon)}{n_0} \ . \eeopt
The amount of communication in this protocol is proportional to the
number of copies of $\rho_{ABE}$. Since $\epsilon$ was arbitrary we
obtain a sequence of protocols (each with communication linear in
the number of copies $\rho^{AB}$) which approaches the rate $R$.
\end{proof}

\section[Proof of Theorem~3.5]{Proof of Theorem~\ref{thm:intr-bound}} \label{app:intrprop}

To show that the intrinsic information is an upper bound on the key
rate, it suffices to verify that it satisfies the requirements of
Theorem~\ref{thm:secrecy-mon}. We start by proving monotonicity of
$I(A:B|E')$ under LOPC operations.

\emph{Monotonicity.} Local operations, i.e., operations on either
Alice's or Bob's side, consist of adding a local ancilla system,
applying a local unitary transformation and removing a local
subsystem. The two first operations leave $I(A:B|E')$ constant. So we
have to show that $I(A:B|E')$ does not increase under partial trace,
i.e. $I(A:B'|E')\leq I(A:B'B''|E') $. This follows immediately from
chain rule and the positivity of the quantum mutual information: \be
\begin{split} I(A:B'|E')&=I(A:B'B''|E')-I(A:B''|E'B') \leq
I(A:B'B''|E').
\end{split}\ee
Public communication from Alice to Bob is the process where a
classical register $C$ is copied to both Bob and Eve. This can be
done in two steps. First, two copies $C'$ and $C''$ of $C$ are
created locally on Alice's side, hence $I(ACC'C'';B|E')=
I(AC;B|E')$. Second, Alice hands over $C'$ to Eve and $C''$ to Bob.
In order to conclude that the conditional mutual information is
non-increasing under public communication, it therefore suffices to
show that $I(AC;BC''|C'E')\leq I(ACC'C'';B|E')$. Writing it out in
terms of entropies, the claim is
\beopt S(ACC'E')+S(BC'C''E')-S(C'E')\leq S(ACC'C''E')+S(BE')-S(E') \eeopt
which is equivalent to
\beopt S(ACE')+S(BCE')-S(CE')\leq S(ACE')+S(BE')-S(E') \eeopt
since $C'$ and $C''$ are copies of $C$. Eliminating the term
$S(ACE')$ we see that the claim is true by strong subadditivity of
von Neumann entropy.

Since the statement holds for arbitrary channels, it also holds for
the intrinsic information.

\paragraph{Asymptotic continuity.}
Let $|| \rho_{ABE}-\sigma_{ABE}||_1\leq \epsilon$. Since the trace
distance is non-increasing under CPTP maps we find $||
\rho_{ABE'}-\sigma_{ABE'}||_1\leq \epsilon$, where
$\rho_{ABE'}=(I_{AB}\ot \Lambda_{E\to \bar E})(\rho_{ABE})$ and
$\sigma_{ABE'}=(I_{AB}\ot \Lambda_{E\to \bar E})(\sigma_{ABE})$. By
the conditional version of Fannes' inequality we find
\beopt |I(A:B|E')_{\rho}-I(A:B|E')_{\sigma}|\leq 8 \epsilon \log d_A +4 H(\epsilon) \ . \eeopt

Since the statement holds for arbitrary channels, it also holds for
the intrinsic information.

\paragraph{Normalisation.} This property can be verified by inserting $\tau_{ABE}$ into the definition of the intrinsic information.

\paragraph{Subadditivity.} We first prove additivity on tensor products for the mutual information:
\begin{align} 
  I(A_1 A_2: B_1 B_2 | E'_1 E'_2) & = I(A_1:B_1|E'_1 E'_2) +
  \underbrace{I(A_1:B_2|E'_1 E'_2 B_1)}_{=0} \\ & \quad
  + I(A_2:B_2|E'_1 E'_2 A_1 ) + \underbrace{I(A_2:B_1|E'_1 E'_2 A_1 B_2)}_{=0} \nonumber \\
  & = I(A_1:B_1|E'_1 ) + I(A_2:B_2|E'_2)
\end{align}
where the last inequality follows by the independence of $\rho_{A_1
  B_1 E'_1}$ and $\rho_{A_2 B_2 E'_2}$. Subadditivity for the
intrinsic information follows from the observation that the infimum in
the definition includes product channels. \qed

\section[Proof of Theorem~3.12]{Proof of Theorem~\ref{prop-upper-separable}} \label{app:boundpr}


The statement $K_D(\rho_{A B} \otimes \rho_E) \leq I(A:B)$ follows
from the intrinsic information bound. The equality condition is a
consequence of~\eqref{eq:lowerbound}.
  
To prove the second part of the theorem, we view $\rho_{AB}$ as the
partial state of a tripartite state \beopt \rho_{ABD}=\sum_i p_i
|i\>_D\<i| \ot \rho_A^{(i)}\ot \rho_B^{(i)} \eeopt where $D$ is a
classical register.  Consider a key distillation protocol for the
state $\rho_{A B}$ with rate $\cR$.  For the map on $n$ copies of the
state, let $C$ be the overall communication and let $A'$ and $B'$
denote the classical keys generated by Alice and Bob, respectively.
The definition of the key rate implies that $I(A':B'|C)/n$ converges
to $\cR$. It is our goal to find an upper bound for $I(A':B'|C)$.

Note that $I(A':B'|CD)=0$ since the only correlations between Alice
and Bob come from $D$ and from communication. Thus from the chain rule
we get
$I(A':B'D|C)= I(A':D|C)$ so that 
\beopt I(A':B'|C) \leq I(A':D|C)
   = I(A'C:D|C) \leq I(A'B'C:D). 
\eeopt
The r.h.s.\ is a lower bound for the LOPC-accessible information of
$n$ copies of the ensemble $\cE$.  However LOPC-accessible information
is additive if its members are separable \cite{hiding-ieee}.  Thus we
obtain \beopt I_{\acc}^{\LOPC}(\{p_i, \rho_A^i\ot \rho_B^i\})\geq {1\over
  n} I(A':B'|C) \to \cR \ .\eeopt Of course $I_{\acc}$ is by definition
not smaller than $I_{\acc}^{\LOPC}$. This concludes the proof since
the above for any protocol. \qed

%
%


\end{document}